\documentclass[preprint,showpacs,aps,11pt]{revtex4}

\usepackage{graphicx}
\usepackage{dcolumn}
\usepackage{amsmath}
\usepackage{amssymb}
\usepackage{equations}

\newcommand{\dotprod}{{\scriptscriptstyle \stackrel{\bullet}{{}}}}

\begin{document}

%\preprint{HEP/123-qed}

\title{Power-Law Behavior of Power Spectra \\
in Low Prandtl Number Rayleigh-B\'{e}nard Convection}

\author{M. R. Paul}
 \email{mpaul@caltech.edu}
\author{M. C. Cross}
\affiliation{Department of Physics, California Institute of Technology 114-36, 
Pasadena, California 91125}

\author{P. F. Fischer}
\affiliation{Mathematics and Computer Science Division, Argonne National Laboratory, 
Argonne, Illinois 60439}

\author{H. S. Greenside}

\affiliation{Department of Computer Science and Department of Physics, 
Duke University, Durham, North Carolina 27706}

\date{\today}% you may specify any date with \date.

\begin{abstract}
The origin of the power-law decay measured in the power spectra of 
low Prandtl number Rayleigh-B\'{e}nard convection near the onset of 
chaos is addressed using long time numerical simulations of the 
three-dimensional Boussinesq equations in cylindrical domains. The 
power-law is found to arise from quasi-discontinuous changes in the 
slope of the time series of the heat transport associated with the 
nucleation of dislocation pairs and roll pinch-off events. For larger 
frequencies, the power spectra decay exponentially as expected for 
time continuous deterministic dynamics.
\end{abstract}


\maketitle

Significant insight into the onset of chaotic dynamics in fluid 
systems, and continuum systems in general, has been gained from 
cryogenic Rayleigh-B\'{e}nard convection 
experiments~\cite{cross:1993,behringer:1985}. Two of the most dramatic 
discoveries were the observation of time dependence almost immediately 
above the onset of convective flow, and the power-law fall-off in 
frequency for the power spectral density derived from time series of 
a global measurement of the temperature difference across the fluid at 
fixed heat flow~\cite{ahlers:1974,ahlers:1978,ahlers:1980}. However, 
these and other important observations remain poorly understood. 
The power-law behavior was unexpected, since bounded deterministic models 
typically show an exponential falloff at high frequency~\cite{frisch:1981}. 
Phenomenological stochastic models were proposed to explain the 
spectra~\cite{greenside:1982}, but no understanding of the origin of the 
ad hoc stochastic driving has followed.

In this paper, we use numerical simulations of the three-dimensional 
Boussinesq equations for the fluid flow and heat transport in the 
cylindrical geometries of the experiments with realistic boundary 
conditions to investigate the power spectrum in more detail. The 
numerical simulations allow us to determine the spatial structure of 
the flow field in the aperiodic dynamics, and the absence of experimental 
or measurement noise provides us with more complete results for the power 
spectra. Our completely deterministic simulations yield results consistent 
with the experimental observations, including a power-law falloff of the 
power spectrum over the range accessible to the experiment. Using 
knowledge of the flow field, we are able to associate this power-law behavior 
with specific events in the dynamics, namely, the creation and annihilation 
of defects in the convection roll structure, which occur on a time scale rapid 
compared with the slow pattern evolution. At higher frequencies, the power 
spectra decay exponentially, consistent with the behavior expected for 
smooth deterministic time evolution. The low amplitude region of the spectra 
was inaccessible experimentally due to the noise floor.

Our simulations in a cylindrical geometry are performed using 
an efficient spectral element algorithm (described in detail elsewhere 
\cite{fischer:1997}). The velocity $\vec{u}$, temperature $T$, 
and pressure $p$, evolve according to the Boussinesq equations,
\begin{eqnarray*}
  {\sigma}^{-1} \left(
  {\partial}_t + \vec{u} \dotprod \vec{\nabla} \right) \vec{u}
  &=&  -\vec{\nabla} p + RT \hat{z} + \nabla^2 \vec{u}  , \\
  \left( {\partial}_t + \vec{u} \dotprod \vec{\nabla} \right) T
  &=& \nabla^2 T  , \\
  \vec{\nabla} \dotprod \vec{u} &=& 0,
\end{eqnarray*}
where $\partial_t$ indicates time differentiation, $\hat{z}$ is a unit vector 
in the vertical direction, $\sigma$ is the Prandtl number, and $R$ is the 
Rayleigh number. The equations are nondimensionalized in the standard manner 
using the layer depth $h$, the vertical diffusion time for heat ${\tau}_v$, and 
the constant temperature difference across the layer $\Delta T$, 
as the length, time, and temperature scales, respectively.  All variables in the 
following discussion are nondimensional using this scaling. The lower and upper 
surfaces ($z=0,1$) are no-slip and are held at constant temperature. The sidewalls 
are no-slip and perfectly conducting~\footnote{The cryogenic 
experiments~\cite{behringer:1982} were bounded by a thin stainless steel lateral 
wall separating the convective layer from a surrounding vacuum resulting 
in the diversion of $\sim 20\%$ of the heat flow suggesting an insulating 
boundary. The room-temperature argon experiments~\cite{croquette:1986,pocheau:1989}, 
on the other hand, were bounded by highly conducting sidewalls. We have performed 
simulations using both lateral boundary conditions and find that the general 
nature of the resulting dynamics is unaffected.}, and the initial conditions 
are small random thermal perturbations of magnitude 0.2 imposed upon an 
otherwise quiescent layer, $\vec{u} = 0$, $T = 0$.

In nearly all cryogenic experiments, for reasons of increased experimental 
resolution, the heat flux across the convection layer, $Q$, and the 
temperature of either the upper or lower surface are held constant while 
measurements of $\Delta T (t)$ are made. These measurements are reported as 
$R(t)/R_c$ or $\Delta T(t)/{\Delta T}_c$, where ${\Delta T}_c$ is the 
temperature difference across the layer and $R_c$ is the Rayleigh number 
at the convective threshold. Theoretical calculations, on the other hand, 
most often consider both the upper and lower surfaces to be held at constant 
temperature and observe the time dependence in $Q(t)$, which can be reported 
as a time series of the Nusselt number $N(t)$ (the normalized heat current 
through the fluid layer). It has been shown experimentally that fixing $Q$ or 
fixing $\Delta T$ does not appear to change the flow dynamics, and the conclusions 
from measurements of $R(t)$ at fixed $N$ or $N(t)$ at fixed $R$ 
will be similar \cite{gao:1984}.

In order to make contact with experiment~\cite{ahlers:1978:supp,ahlers:1980,gao:1984} 
we focus our discussion on simulations with aspect ratio $\Gamma=4.72$ ($\Gamma = r/h$, 
$r$ is the radius), $\sigma=0.78$ (experimental fluid was nonsuperfluid He${}^4$), 
and constant $\Delta T$. A key result of the experiments was the observation 
that the power spectrum, $P(\nu)$, of measured $R(t)$ values exhibited the 
power-law behavior, $P(\nu) \sim {\nu}^{-n}, n = -4.0 \pm 0.2$ over the 
frequency range $0.5 \lesssim \nu \lesssim 9$~\cite{ahlers:1978:supp} (results 
were reported for $\epsilon=3.62$, where $\epsilon = (R-R_c)/R_c$ is the 
reduced Rayleigh number).

\begin{figure}
[tbh]
\begin{center}
\includegraphics[width=3.0in]{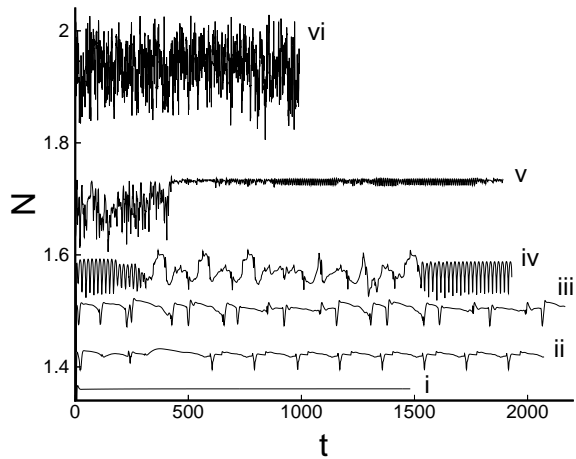}
\end{center}
\caption{Plots of the dimensionless heat transport N(t) for cases (i-vi) for 
reduced Rayleigh number $\epsilon = 0.557,0.614,0.8,1.0,1.5$, 
and $3.0$, respectively. For cases (i-v), $\Delta t =0.01$, and for case~(vi), 
$\Delta t = 0.005$ ($\Delta t$ is the time step).}
\label{fig:numany}
\end{figure}

Six representative time series $N(t)$ from our simulations 
are shown in Fig.~\ref{fig:numany}. In terms of the horizontal diffusion 
time for heat ${\tau}_h$ (${\tau}_h={\Gamma}^2 {\tau}_v$), 
the simulation times are $t_f \approx 100 {\tau}_h$ 
($t_f \approx 50 {\tau}_h$ for case~(vi)), which is comparable with the 
longest experiments, $t_f \approx 65 {\tau}_h$ (with one long run for 
$t_f \approx 135 {\tau}_h$)~\cite{ahlers:1980}. This is considerably 
longer than $\Gamma {\tau}_h$, which has been suggested as the earliest 
time scale for the flowfield to reach equilibrium~\cite{cross:1984}. 
However, as discussed below, we find that the dynamics can occur on 
even longer time scales. The simulated time-averaged values of $N-1$ 
are within $5.5\%$ of the experimental values given by 
$N-1=1.034 \beta + 0.981 \beta^3 - 0.866 \beta^5$, 
$\beta=1-R_c/R$~\cite{ahlers:1974}. Spatial and temporal resolution 
studies have been performed to ensure the accuracy of the calculated values 
of $N(t)$ for the chosen simulation parameters.

\begin{figure}
[tbh]
\begin{center}
\includegraphics[width=3.0in]{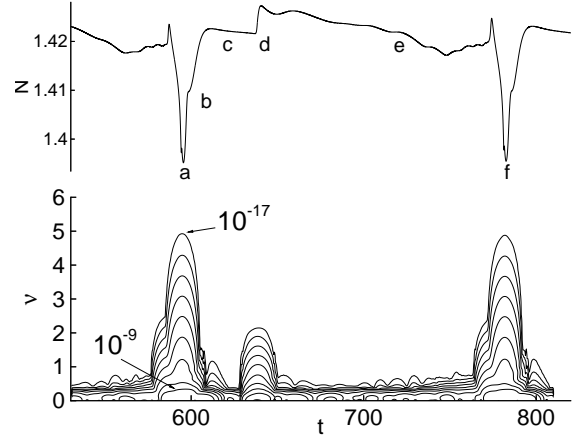}
\end{center}
\caption{Time series $N(t)$ (top) and corresponding spectrogram 
(bottom) for one period of case~(ii). The labels a-f represent particular 
moments in the evolution of the pattern and are discussed in the text. 
The spectrogram displays 9 orders of magnitude of the power, $P(\nu)$, with 
the smallest and largest contours labeled; the remaining contours each differ 
by a factor of 10. The spectrogram was calculated using a sliding Hann window 
of width $\Delta t = 20.48$ and linearly detrended overlapping segments 
(segments overlap by $t=20.0$).}
\label{fig:spec}
\end{figure}

\begin{figure}
[tbh]
\begin{center}
\includegraphics[width=3.0in,height=1.5in]{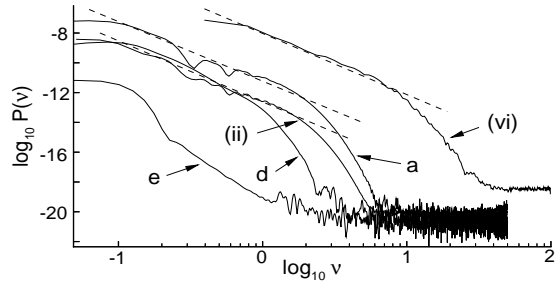}
\end{center}
\caption{Windowed power spectra.  The power spectra labeled (a), (d), 
and (e) are vertical slices of the spectrogram taken at representative times for 
case~(ii) at dislocation nucleation, dislocation annihilation, and dislocation 
glide, respectively. The curves labeled (ii) and (vi) are the average of the 
windowed power spectra using the entire spectrogram for cases (ii) and (vi). 
The dashed lines represent $P(\nu) \sim \nu^{-4}$.}
\label{fig:psd_compare}
\end{figure}

\begin{figure}
[tbh]
\begin{center}
\includegraphics[width=2.5in,height=2.5in]{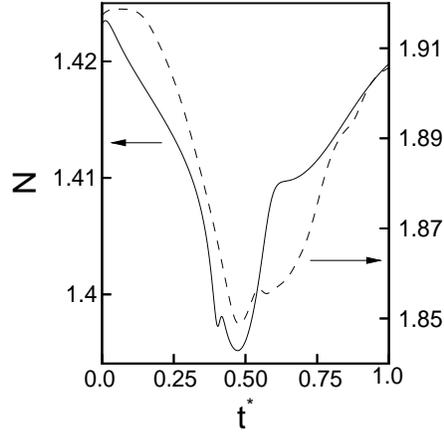}
\end{center}
\caption{A closeup of the time series $N(t)$, illustrating the 
signature of a nucleation of a dislocation pair for case~(ii) 
(solid line), and the signature of a roll pinch-off for case~(vi) 
(dashed line). Time $t^*$ is measured as $t^* = (t - t_i)/\Delta t$; 
$t_i$ denotes when the event begins, and $\Delta t$ is the duration 
of the event. For the events shown, $t_i=587,270.5$ and $\Delta t=18,1.5$ 
for cases (ii) and (vi), respectively.}
\label{fig:nu_compare}
\end{figure}

We now consider the periodic time series in more detail, case~(ii) 
in Fig.~\ref{fig:numany}. To determine the influence of the pattern 
dynamics on the power spectrum, we used a sliding window in time to 
calculate successive time-localized power spectra (a spectrogram) as 
shown in Fig.~\ref{fig:spec}b. Most of the power in the spectra can 
be attributed to nucleation of dislocation pairs (and, to a lesser 
extent dislocation annihilation). A plot of $P(\nu)$ at a particular 
time (a vertical slice of Fig.~\ref{fig:spec}b) yields the windowed 
power spectrum centered about that point in time in Fig.~\ref{fig:spec}a. 
Figure~\ref{fig:psd_compare} shows three such power spectra from the 
spectrogram evaluated with windows centered on (a), (d), and (e) 
corresponding to dislocation nucleation, annihilation, and glide, 
respectively. The local power spectrum centered 
on the nucleation of a dislocation pair generates 
a power-law region of significant magnitude; the local power spectrum 
centered on the dislocation annihilation generates a power-law region 
that is a factor of 10 smaller in magnitude; whereas the local power spectrum 
calculated during dislocation glide falls off more rapidly with $\nu$ and 
does not make a significant contribution. The origin of the power-law is the 
time signature of the nucleation of a dislocation pair manifested in 
the quasi-discontinuous slope of $N$ during the rapid excursion to 
a state of decreased heat transfer (an individual event is shown 
in Fig.~\ref{fig:nu_compare}). Note that a triangular feature 
with discontinuous changes in the slope yields a $\nu^{-4}$ asymptotic 
behavior in the power spectrum.

The power spectra of the remaining time series in Fig.~\ref{fig:numany} 
yield a power-law region $P(\nu) \sim {\nu}^{-4}$ because the 
dynamics (periodic, quasiperiodic, and chaotic) are dominated by the 
nucleation of dislocation pairs and roll pinch-off events. The differences 
between the chaotic and periodic dynamics are apparent in the low-frequency 
region and are more apparent on a log-linear plot of $P(\nu)$. 
For example, Fig.~\ref{fig:nu_compare} shows the similarity between $N(t)$ for 
the periodic simulation, case~(ii), and the chaotic simulation, case~(vi). 
The dynamics in the chaotic state are much more complicated; however, they are 
dominated by the roll pinch-off events that maintain the characteristic 
quasi-discontinuous slope of $N$, yielding a power-law region in the power 
spectrum. A comparison of the power spectra for the periodic and chaotic time 
series is shown in Fig.~\ref{fig:psd_compare}.

\begin{figure}
[tbh]
\begin{center}
\includegraphics[width=3.0in,height=1.5in]{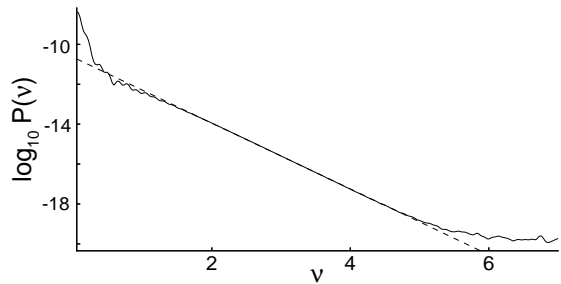}
\end{center}
\caption{The power spectrum, $\left< P(\nu) \right>$, for case~(ii) on a 
log-linear scale to illustrate region of exponential decay. The slope 
of the dashed line is -3.8; the crossover to exponential decay occurs 
at $\nu \approx 1.5$.}
\label{fig:psdexpo}
\end{figure}

The average of the windowed power spectra of Fig.~\ref{fig:psd_compare} 
eventually exhibit an exponential decay, which continues until reaching 
the spectrum floor; this is shown for case~(ii) in Fig.~\ref{fig:psdexpo}. 
Exponential decay in the power spectra at high frequency is expected for 
bounded smooth deterministic dynamics~\cite{frisch:1981}. The exponential 
decay in the power spectra was not detected in experiment due to the 
presence of instrumental noise which masked the small scale region.

\begin{figure}
[tbh]
\begin{center}
\includegraphics[width=3.0in]{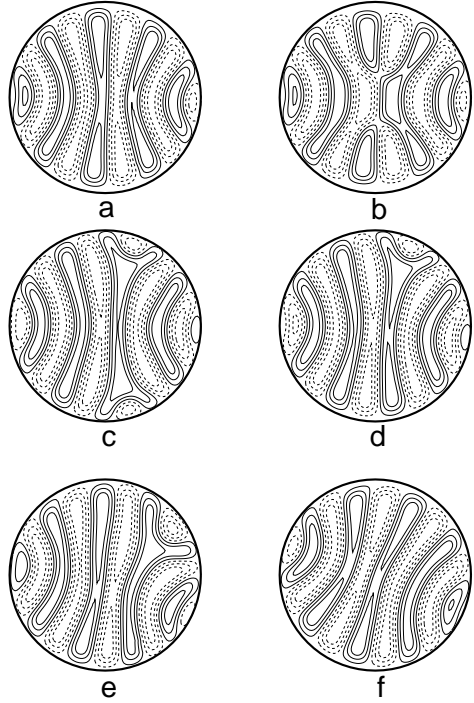}
\end{center}
\caption{Flow visualization showing contours of the thermal perturbation 
at the mid-depth, (6 evenly spaced contours; $-0.2 \leq \delta T \leq 0.2$, 
negative values are dashed lines, and positive values are solid lines) 
for case~(ii). Panels a-f are for $t = 600,605,630,650,735,785$. The 
dislocations glide to the right; during the next period, the dislocations glide to 
the left, as can already be discerned in (f) by the bias in the roll compression. 
This left and right alternation continues for the entire simulation.}
\label{fig:snapshots}
\end{figure}

In the cryogenic experiments, flow visualization was not possible leaving 
the precise details  of the underlying pattern uncertain. With this in 
mind, we briefly discuss the dynamics represented in 
Fig.~\ref{fig:numany}. Case~(i) illustrates a time-independent Pan-Am 
pattern similar to panel (a) of Fig.~\ref{fig:snapshots}. Case~(ii) is 
periodic with period $t=8.4{\tau}_h$ (note the initial transient lasting 
$27 {\tau}_h$); the dynamics of one period are illustrated 
in Figs.~\ref{fig:spec} and~\ref{fig:snapshots}. Figure~\ref{fig:snapshots} 
displays the pattern at six different instances in time corresponding to 
the events labeled in Fig.~\ref{fig:spec}a. Initially there is a Pan-Am 
pattern with two opposing wall foci causing roll compression (a), 
eventually nucleating a dislocation pair in the center of the domain (b). 
The dislocations quickly climb to the wall (c), at which point 
they both begin to glide slowly toward the same wall focus. However, 
the lower dislocation is annihilated at the sidewall (d), and the remaining 
dislocation continues to glide slowly into the wall focus, where it is annihilated 
(e). A Pan-Am pattern again forms (f), and finally the process repeats. This 
is in general agreement with flow visualizations from related room-temperature 
argon-based experiments~\cite{croquette:1986,pocheau:1989}. 
Case~(iii) may be periodic on a long time scale of $t \approx 40 {\tau}_h$; 
the duration of the simulation is inadequate to be conclusive. Case~(iv) 
illustrates a chaotic burst of duration $t \approx 54 {\tau}_h$ bounded 
by periodic dynamics with a period of $t \approx 17$. Again the simulation 
duration is inadequate to determine whether this is a transient state or 
whether the chaotic bursting will repeat. Case~(v) shows an initial chaotic transient  
that makes a transition at $t \approx 18 {\tau}_h$ to a very complicated 
quasiperiodic state where the central roll pair is pinned by the dynamic motion 
of two opposing disclinations. The dominant mode in the quasiperiodic state 
has a time scale $t \approx 8$. Case~(vi) illustrates chaotic dynamics.

We have also performed simulations for the $\Gamma=4.72$ cylindrical 
domain with insulating lateral boundary conditions, in addition to simulations 
in a $\Gamma=7.66$ domain ($\sigma=0.69$ for argon) for both 
conducting and insulating lateral boundaries. Considering these 
additional results, we maintain our conclusions concerning the origin 
of the power-law.

This work represents a joint computational and theoretical 
effort to further our quantitative understanding of complex 
dynamics in spatially extended nonequlibrium systems. An 
important link missing in nearly all theoretical work to 
date has been a quantitative comparison with experiment. Our 
results demonstrate that this quantitative comparison with 
experiment is now possible. We plan to use this 
approach to investigate spatiotemporal chaos in larger 
aspect ratio systems.

This research was supported by the U.S.~Department of Energy, 
Grant DE-FT02-98ER14892, and the Mathematical, Information, 
and Computational Sciences Division subprogram of the 
Office of Advanced Scientific Computing Research, 
U.S.~Department of Energy, under Contract W-31-109-Eng-38. 
We also acknowledge the Caltech Center for Advanced 
Computing Research and the North Carolina Supercomputing Center.

%\bibstyle{prsty}  note that the bibstyle is set to prsty in aps.rtx and
%                 for some reason is not read in here.
%\newpage
%\bibliography{plaw}% Produces the bibliography via BibTeX.

%%%%%%%%%%%%%%%%%%%%%%%%%%%%%%%%%%%%%%%%%%%%%%%%%%%%%%%%%%%%%%%%%%%%%%%%%%%%%
%%%%%%%%%%%%%%%%%%%%%%%%%%%%%%%%%%%%%%%%%%%%%%%%%%%%%%%%%%%%%%%%%%%%%%%%%%%%%

%%%%%%%%%%%%%%%%%%%%%%%%%%%%%%%%%%%%%%%%%%%%%%%%%%%%%%%%%%%%%%%%%%%%%%%%%%%%%

%%%%%%%%%%%%%%%%%%%%%%%%%%%%%%%%%%%%%%%%%%%%%%%%%%%%%%%%%%%%%%%%%%%%%%%%%%%%%

%%%%%%%%%%%%%%%%%%%%%%%%%%%%%%%%%%%%%%%%%%%%%%%%%%%%%%%%%%%%%%%%%%%%%%%%%%%%%
%%%%%%%%%%%%%%%%%%%%%%%%%%%%%%%%%%%%%%%%%%%%%%%%%%%%%%%%%%%%%%%%%%%%%%%%%%%%%

%%%%%%%%%%%%%%%%%%%%%%%%%%%%%%%%%%%%%%%%%%%%%%%%%%%%%%%%%%%%%%%%%%%%%%%%%%%%%
%%%%%%%%%%%%%%%%%%%%%%%%%%%%%%%%%%%%%%%%%%%%%%%%%%%%%%%%%%%%%%%%%%%%%%%%%%%%%

\end{document}